# Printed Graphene Circuits**

By *Jian-Hao Chen*, *Masa Ishigami*, *Chaun Jang*, *Daniel R. Hines*, *Michael S. Fuhrer*, and *Ellen D. Williams**

A single layer of graphite, graphene,[1, 2] is a truly 2-dimensional semi-metallic material composed of only one atomic layer of carbon atoms. Graphene's peculiar band structure suppresses carrier backscattering, leading to extremely high carrier mobility.[2] Narrow graphene ribbons are predicted to have a semiconducting energy gap tunable by width,[3] indicating a path to device fabrication. In addition, because graphene is only one atom in thickness, transport properties are expected to be sensitively influenced by atomic scale defects, adsorbates,[4, 5] local electronic environment, and mechanical deformations; consequently, graphene is a promising sensor material. To date, graphene has been obtained by only two methods: mechanical exfoliation of graphite on $SiO_2/Si$[1] or thermal graphitization of a silicon carbide (SiC) surface.[2] In each case, the substrate strongly influences the graphene properties; charge defects in $SiO_2$ are thought to limit the mobility, and strong interaction with SiC introduces a large charge density. Furthermore, the substrate can limit the graphene device possibilities; gating of devices on SiC is difficult, and on $SiO_2/Si$ the presence of a conducting backplane (also used as the gate) precludes high-frequency device operation. In this paper, we report the transfer of graphene from one substrate to another to realize flexible, transparent graphene devices with high field effect mobility. This represents the ultimate extension of printing technology to a single atomic layer.

We employ the transfer printing method[6, 7] to transfer graphene between $SiO_2/Si$ and plastic substrates, as well as to assemble the gate dielectric, and source, drain, and gate electrodes, forming a complete graphene field-effect transistor with local gate on a flexible, transparent substrate. Transfer printing enables device component fabrication and assembly to be performed separately, and has found wide application in printed circuits and flexible electronics research.[7-10] By properly tuning the adhesion of the printed material to the original and target substrate,[7] our technique can in principle enable the transfer of graphene to *any* substrate, thus greatly expanding the possible applications of this material.

Figure 1

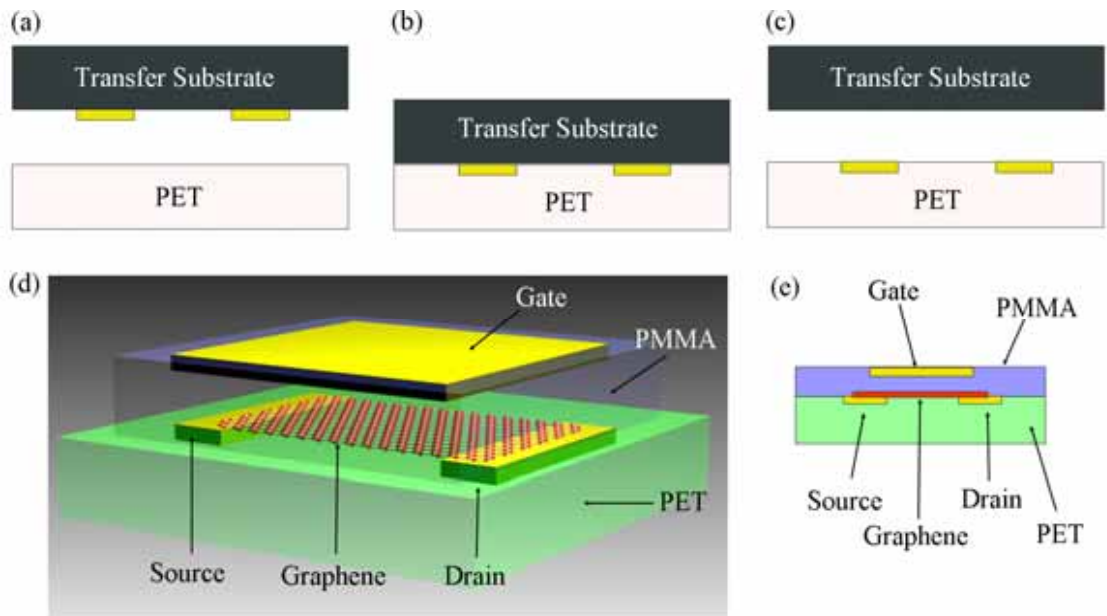

Figure 1 (a)-(c) Printing procedure used to print a feature layer. (a) The desired features, e.g. two gold electrodes, are predefined on the transfer substrate. (b) The transfer substrate is brought into contact with the plastic substrate at an elevated temperature and high pressure. Temperature and pressure are optimized to ensure successful transfers. (c) The transfer substrate is removed from the plastic substrate, leaving the features embedded in the plastic substrate. The process may be repeated to assemble additional components. (d) The 3D schematic and (e) the cross sectional view of the completed graphene device, not drawn to scale.

Fig. 1a-c depicts the basic process required to print a patterned layer of material from one substrate (the transfer substrate) over to a second substrate (a PET plastic substrate). The devices require three process steps performed sequentially to assemble (1) source-drain electrodes, (2) graphene, and (3) gate electrode/dielectric. First, photolithography is used to prepare 30 nm thick Au source and drain electrodes on a silicon wafer with an oxidized surface ($SiO_2$/Si). The electrodes are then transferred onto the PET substrate as described elsewhere.[6, 7] Then, single- and few-layer graphene is obtained from Kish graphite by mechanical exfoliation[1] on 300 nm thick thermally-grown silicon dioxide on silicon substrates, and its thickness and morphology

characterized by atomic force microscopy. Mechanical exfoliation yields atomically-clean graphene sheets[11, 12] and our AFM images also indicate that the graphene sheet is free of nanometer-scale contaminants. In addition, chemical contamination caused by exposure to photoresist and lift-off chemicals is avoided in this process. The desired graphene sheet is printed at 170 °C at 500 psi from the $SiO_2$/Si substrate to the source-drain electrode assembly on PET. Under these conditions, the PET substrate is above its glass transition temperature, and can conform to the transfer substrate morphology.[7] Finally, the gate assembly consisting of a photolithographically patterned 100 nm Au gate electrode and a 600 nm thick poly(methyl methacrylate) (PMMA) gate dielectric is prepared on $SiO_2$/Si and transfer printed onto the device substrate at 175 °C at 500 psi. Each subsequent layer is aligned optically to the pre-existing features. Fig. 1d shows the schematic of a completed device. An advantage of this method is that it exposes graphene to no chemicals used in conventional lithography processes, by which most of the graphene devices on silicon dioxide are fabricated. Lithography processes have been found to leave residue on the device[12] and might negatively influence transport properties.

The printing process is successful in transferring graphene materials, ranging from monolayer sheets to bulk graphite, from the silicon dioxide substrate to PET and Au. Fig. 2a shows an optical microscopy image of a graphite film with thicknesses from monolayer to multilayer on a silicon dioxide substrate. Fig. 2b shows the graphene material printed to the source-drain electrodes on PET (the image is reversed to aid comparison to Fig. 2a). By comparison of Figs. 2a and 2b, it is clear that the conduction from source to drain electrode takes place through the portions labeled "monolayer" and "bilayer" in Fig. 2a, in series. (As a visual aid, red dotted lines have been added to Fig. 2a as an indicator of the location of the edges of the source-drain electrodes (separated by 6 μm) with respect to the graphene before printing.) The thickness of the monolayer portion is confirmed by atomic force microscopy (AFM) before transfer printing as shown in Figs 2c-d. Fig. 2c is an AFM micrograph acquired in the boxed region indicated in Fig. 2a, which shows the functioning monolayer portion with another monolayer lying across it. The red box in Fig. 2c shows an area where the top layer steps down from the functioning layer to the substrate, and the step height here is the thickness of the functioning layer. Fig. 2d shows the height histogram of the area inside the red box in Fig. 2c. Fitting the histogram by two Gaussian peaks gives an estimate of the thickness of the monolayer portion to be 3.95±0.09Å, which confirms that the functioning material is single layer graphene.[12]

Figure 2

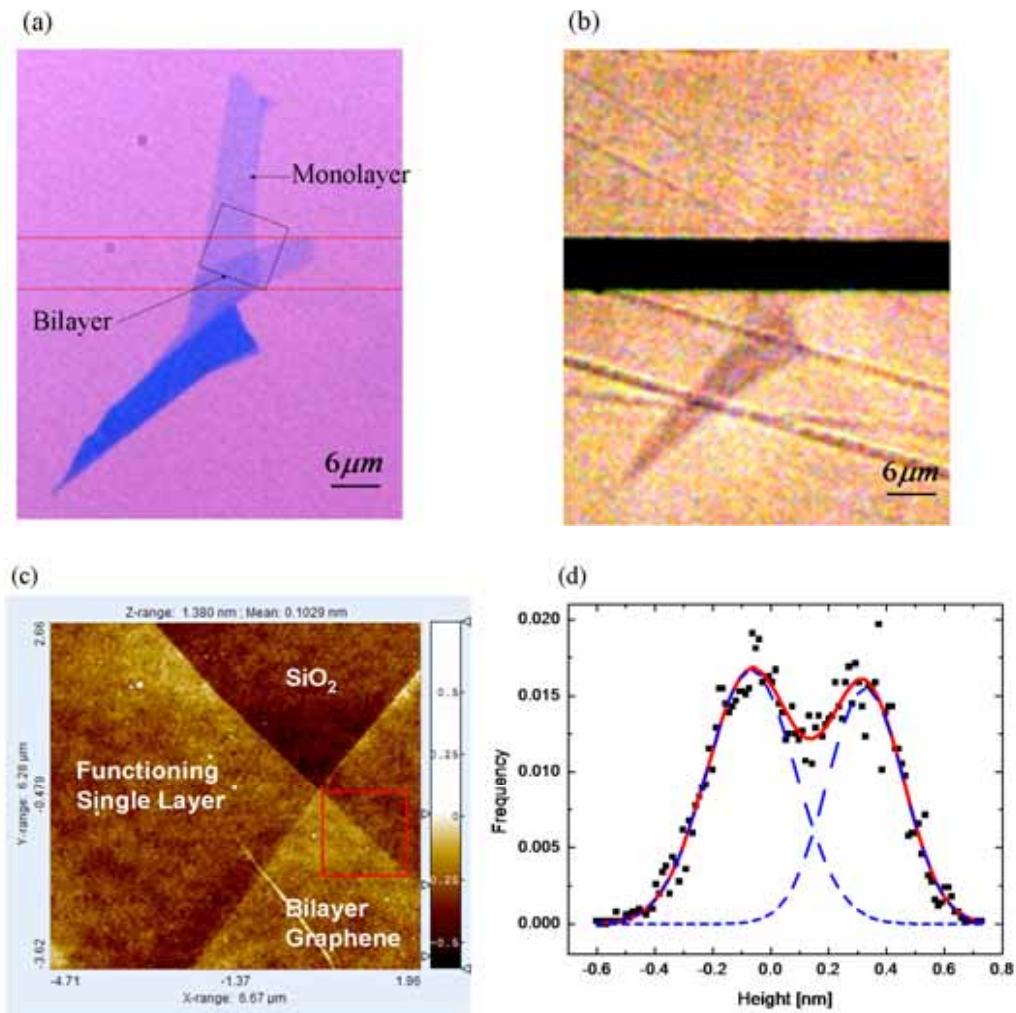

Figure 2. (a) Optical microscopy image of a mixed monolayer and multilayer graphene material on silicon dioxide substrate. (b) Optical microscopy image of the same graphene sample transfer printed onto the source/drain electrode assembly (dark area is PET, yellow areas are Au electrodes). The Au source-drain electrodes are bridged by graphene composed of a single-layer portion and a bilayer portion. Note: (b) is left-right reversed to aid comparison to (a). (c) Atomic force micrograph of the over lapping area of the sample in (a), used to determine the number of graphene layers. (d) Histogram of the selected area (area inside the red box) is fitted by two Gaussian peaks. The height

difference between the two peaks is 3.95±0.09Å, which indicates that the functioning material is single layer graphene.

After transfer no graphene is observed in optical images on the silicon dioxide substrate; this indicates that graphene adheres more strongly to PET and Au than to the original silicon dioxide substrate, and the interlayer coupling strength of graphite is stronger than its adhesion to the silicon dioxide surface. The presence of the Au source-drain electrodes is not necessary for transfer of graphene materials from silicon dioxide substrates to PET; graphene materials can be transferred to bare PET, as suggested by simulations.[13] Graphene materials are barely visible once transferred onto PET as seen in Fig.2b, and can only just be discerned on the source-drain electrodes. Graphene is nearly completely transparent at visible wavelengths.

Figure 3

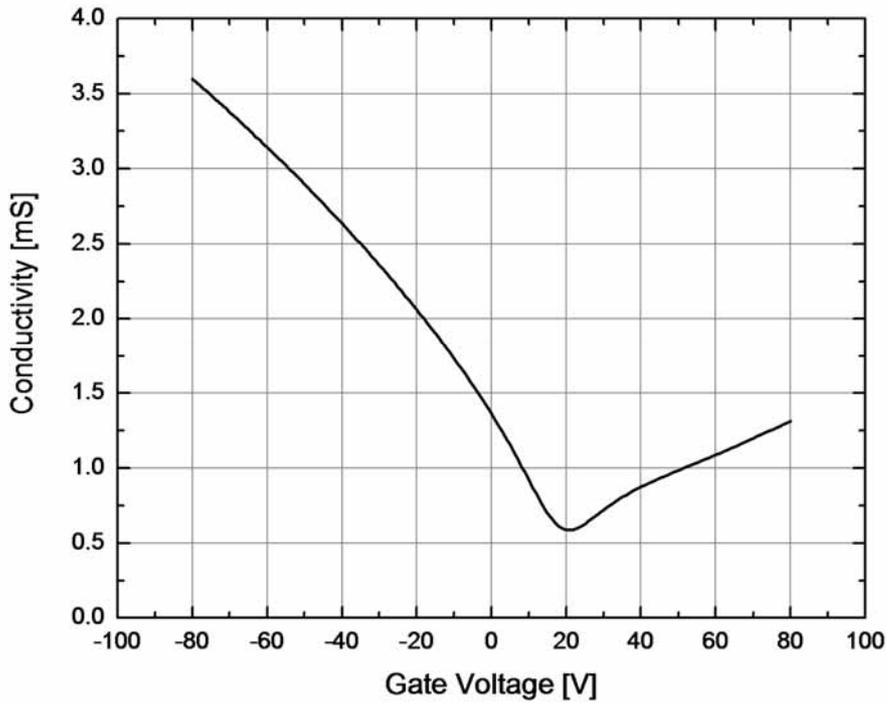

Figure 3. Conductivity as a function of gate voltage for the device in Fig. 2b measured at 297 K. The minimum conductivity is about 0.6 mS or ≈$8G_0$ at the Dirac neutral point $V_D = 21V$, where $G_0 = 2e^2/h$ is the quantum of conductance. The capacitance of PMMA dielectric used is 4.4 nF/cm². 21 V applied across the PMMA dielectric induces the same amount of charge density as 8V across

300nm silicon dioxide dielectric. Source-drain bias of 10 mV was applied while acquiring the above data.

Measurement of the transport properties is important to assess the usefulness of the transfer printing process. Figure 3a shows the room temperature conductivity[14] as a function of gate voltage $\sigma(V_g)$ of the "printed" device shown in Fig. 2b. As seen in Figs. 2a and 2b, this device consists of two portions (monolayer and bilayer) in series. As graphene sheets are semi-metals with linearly vanishing electronic densities of states at the charge-neutral point, the applied gate voltage modifies the conductivity. The slope of the linear portion of the transfer curve is used to calculate the field effect mobility, $\mu = \frac{1}{c_g}\frac{d\sigma}{dV_g}$, where $c_g$ is the gate capacitance per unit area (4.4 nF/cm$^2$). This particular device shows a maximum field effect mobility of $1.0 \times 10^4$ cm$^2$/Vs for holes and $4 \times 10^3$ cm$^2$/Vs for electrons. Another device composed solely of a monolayer material showed similar field effect mobilities. These values are comparable to the best field effect mobilities measured for graphene devices on SiO$_2$ at room temperature, for example $2 \times 10^3 - 5 \times 10^3$ cm$^2$/Vs reported by Novoselov et al.,[1] and $2 \times 10^3 - 2 \times 10^4$ cm$^2$/Vs reported by Tan et al.,[15] suggesting that the transfer method does not damage the graphene and no chemical bonding was established between graphene and plastic substrates.

The minimum conductivity for the "printed" graphene device, shown in Fig. 3 is approximately 0.6 mS or $\approx 8 G_0$, where $G_0 = 2e^2/h$ is the quantum of conductance. The minimum conductivity reported for monolayer[16] and bilayer[17] graphene-based devices is often near $2G_0$ (but may be higher in clean samples[18]). The high value of the minimum conductivity for the printed devices indicates that the contact resistance is small in the transfer-printed devices. Overall, the results show that transfer printing graphene can yield electronic devices equaling the performance of the conventional silicon dioxide-supported devices.

The Dirac neutral point of the printed device (see Fig. 3) is about 21V (a second printed single layer graphene device showed the same shift), which corresponds to net positive charge density of 5.8 $\times 10^{11}$ cm$^{-2}$. One possible explanation is that this shift originates from excess positive trapped charge in the polymer substrate.[19-22] The same amount of charge density would be induced by applying 8V of gate voltage on 300nm silicon dioxide dielectric. A shift of this magnitude is not uncommon in graphene devices on silicon dioxide,[23] but smaller values have been reported.[24] For comparison, the density of charge traps has been reported to be $2 \times 10^{11}$ cm$^{-2}$ in PET,[20, 22] $5 \times 10^{8}$ cm$^{-2}$ in PMMA,[19, 21] and $5 \times 10^{11}$ cm$^{-2}$ in thermally grown silicon dioxide.[25] If the Dirac point shift is predominately determined by trapped charge, these observations would suggest that the PET/PMMA sandwich creates an excess of positive trapped charge and a net charge density comparable to the best observed devices on SiO$_2$. Alternatively, other mechanisms such as a surface dipole moment, work-function difference between graphene and gate, or chemical doping may also be involved.

Finally, electronic[5, 18, 26] and structural[12] disorder imposed by the substrate, are expected to determine the graphene transport properties, including the mobility, minimum conductivity, and the shift of the Dirac point. The PET/PMMA sandwich substrates in the printed devices nominally[19-22, 25] have net trap densities similar to silicon dioxide substrates. The RMS roughness of the substrate is larger for the PET substrates (1.2 nm in a 5 μm x 5 μm area) than for the silicon dioxide substrates (0.25 nm in a 5 μm x 5 μm area). There are two anomalous features in the transport measurements. First, the minimum conductivity is unusually high at $8G_0$ even for the solely monolayer printed devices. This indicates that the reported universality[16] of the minimum conductivity at $2G_0$ is not correct, and the $2G_0$ value may be specific only to certain silicon dioxide-supported devices. In addition, the devices on the plastic substrates always have higher hole mobility (e.g., they do not have electron-hole symmetry). Such asymmetry has not been reported previously for graphene devices. The present transport theory,[18, 26-28] which focuses on short range or long range scatterers to describe the transport properties, is incapable of explaining such a large asymmetry. The roughness of the PET substrate and the observed high mobility of the printed devices suggest either that the graphene morphology plays little role in determining the transport properties, or that the graphene sheet does not closely conform to the underlying PET morphology. Further work correlating the transport characteristics with systematic variation of substrate charge density and roughness is needed to identify the mechanism behind the differences.

The transfer-printed devices represent the first realization of a local electrostatic gate on graphene-on-insulator. Local gating enables the reduction of gate-source capacitance, which is necessary for high-frequency device operation. Local gating can also be used to explore p-n junctions in graphene, which are predicted to have unusual properties,[29-31] and may form the basis of new bipolar transistor devices.[29] In addition, graphene may represent the ultimate transparent electrode; the resistivity of our graphene at high gate voltage is less than 300 Ω/square, while graphene on PET is so transparent as to be nearly undetectable in the optical microscope.

In conclusion, we have fabricated transparent electronic devices based on graphene materials with thickness down to a single atomic layer by the transfer printing method. The resulting printed graphene devices retain high field effect mobility and have low contact resistance. The results show that the transfer printing method is capable of high-quality transfer of graphene materials from silicon dioxide substrates, and the method thus will have wide applications in manipulating and delivering graphene materials to desired substrate and device geometries. Since the method is purely additive, it exposes graphene (or other functional materials) to no chemical preparation or lithographic steps, providing greater experimental control over device environment for reproducibility and for studies of fundamental transport mechanisms. Finally, the transport properties of the graphene devices on the PET substrate demonstrate the non-universality of minimum conductivity and the incompleteness of the current transport theory.

*Experimental*

*Sample Preparation*: Graphene samples are obtained from Kish graphite by mechanical exfoliation[1] on 300nm thermally-grown silicon dioxide on silicon substrates. Their thickness and morphology are characterized by an atomic force microscope (Digital Instruments (R) IIIa) in the ambient environment.

*Device Fabrication*: The devices require three process steps performed sequentially to assemble (1) source-drain electrodes, (2) graphene, and (3) gate electrode/dielectric. First, photolithography is used to prepare 30 nm thick Au source and drain electrodes on a silicon wafer with an oxidized surface ($SiO_2$/Si). The electrodes are then transferred onto the PET substrate as described elsewhere.[6, 7] The desired graphene sheet is printed at 170 °C at 500 psi from the silicon dioxide substrate to the source-drain electrode assembly on PET. Finally, the gate assembly consisting of a photolithographically patterned 100 nm Au gate electrode and a 600 nm thick poly(methyl methacrylate) (PMMA) gate dielectric is prepared on $SiO_2$/Si and transfer printed onto the device substrate at 175 °C at 500 psi. Each subsequent layer is optically aligned to the pre-existing features.

*Electrical Characterization*: The transfer curves are measured using a probe station in the ambient environment. Source-drain bias voltages of 5mV, 10mV and 20mV are used.

[*] Prof. E. D. Williams, J.H. Chen, Dr. M. Ishigami, C. Jang, D. R. Hines, Prof. M. S. Fuhrer
  Physics Department, University of Maryland, College Park, MD 20742, U.S.A.
  Email: edw@umd.edu

  Prof. E. D. Williams, J.H. Chen, C. Jang, Prof. M. S. Fuhrer
  Materials Research Science and Engineering Center, University of Maryland, College Park, MD 20742, U.S.A.

  C. Jang, Prof. M. S. Fuhrer
  Center for Superconductivity Research, University of Maryland, College Park, MD 20742, U.S.A.
  Prof. E. D. Williams, D. R. Hines
  Laboratory for Physical Sciences, College Park, MD 20742, U.S.A.

  Prof. E. D. Williams
  Institute for Physical Science and Technology, University of Maryland, College Park, MD 20742, U.S.A.

[**]This work has been supported by the Intelligence Community Postdoctoral Fellowship program, the Laboratory for Physical Sciences, the U.S. Office of Naval Research grant no. N000140610882, and NSF grant no. CCF-06-34321. The UMD-MRSEC Shared Equipment Facilities were used in this research.

## References

[1]   K. S. Novoselov, D. Jiang, F. Schedin, T. J. Booth, V. V. Khotkevich, S. V. Morozov, A. K. Geim, *Proc. Natl. Acad. Sci. USA* **2005**, *102*, 10341.
[2]   C. Berger, S. Zhimin, L. Xuebin, W. Xiaosong, N. Brown, C. Naud, D. Mayou, T. Li, J. Hass, A. N. Marchenkov, E. H. Conrad, P. N. First, W. A. de Heer, *Science* **2006**, *312*, 1191.

[3]     V. Barone, O. Hod, G. E. Scuseria, *Nano Lett.* **2006**, *6*, 2748.
[4]     F. Schedin, A. K. Geim, S. V. Morozov, E. W. Hill, P. Blake, M. I. Katsnelson, K. S. Novoselov, *Nat. Mater.* **2007**, *advanced online publication*.
[5]     E. H. Hwang, S. Adam, S. Das Sarma, A. K. Geim, *Preprint at* <http://lanl.arxiv.org/abs/cond-mat/0610834> **2006**.
[6]     D. R. Hines, S. Mezhenny, M. Breban, E. D. Williams, V. W. Ballarotto, G. Esen, A. Southard, M. S. Fuhrer, *Appl. Phys. Lett.* **2005**, *86*, 163101.
[7]     D. R. Hines, V. W. Ballarotto, E. D. Williams, Y. Shao, S. A. Solin, *J. Appl. Phys.* **2007**, *101*, 024503.
[8]     S. R. Forrest, *Nature* **2004**, *428*, 911.
[9]     J.-H. Ahn, H.-S. Kim, K. J. Lee, S. Jeon, S. J. Kang, Y. Sun, R. G. Nuzzo, J. A. Rogers, *Science* **2006**, *314*, 1754.
[10]    S. Yugang, M. Etienne, A. R. John, K. Hoon-Sik, K. Seiyon, C. Guang, A. Ilesanmi, D. Ross, C. Rebecca, T. Alan, *Appl. Phys. Lett.* **2006**, *88*, 183509.
[11]    E. Stolyarova, K. T. Rim, S. Ryu, J. Maulzsch, P. Kim, L. E. Brus, M. S. Hybertsen, G. W. Flynn, *Proc. Natl. Acad. Sci. USA* **2007**, *104*, 9209.
[12]    M. Ishigami, J. H. Chen, W. G. Cullen, M. S. Fuhrer, E. D. Williams, *Nano Lett.* **2007**, *7*, 1643.
[13]    D. J. Henry, C. A. Lukey, E. Evans, I. Yarovsky, *Mol. Simul.* **2005**, *31*, 449.
[14]    *We observe that Id vs. Vd is linear with Vd < 20 mV used for our experiment.*
[15]    Y.-W. Tan, Y. Zhang, K. Bolotin, Y. Zhao, S. Adam, E. H. Hwang, S. Das Sarma, H. L. Stormer, P. Kim, *Preprint at* <http://xxx.lanl.gov/abs/0707.1807> **2007**.
[16]    K. S. Novoselov, A. K. Geim, S. V. Morozov, D. Jiang, M. I. Katsnelson, I. M. Grigorieva, S. V. Dubonos, A. A. Firsov, *Nature* **2005**, *438*, 197.
[17]    K. S. Novoselov, E. McCann, S. V. Morozov, V. I. Fal'ko, M. I. Katsnelson, U. Zeitler, D. Jiang, F. Schedin, A. K. Geim, *Nat. Phys.* **2006**, *2*, 177.
[18]    S. Adam, E. H. Hwang, V. M. Galitski, S. Das Sarma, *Preprint at* <http://arxiv.org/abs/0705.1540> **2007**.
[19]    T. Maeno, T. Futami, H. Kushibe, T. Takada, *J. Appl. Phys.* **1989**, *65*, 1147.
[20]    E. R. Neagu, J. N. Marat-Mendes, R. M. Neagu, D. K. Das-Gupta, *J. Appl. Phys.* **1999**, *85*, 2330.
[21]    D. Sadovnichii, A. Tuytnev, Y. Milekhin, *High Energy Chem.* **2005**, *39*, 148.
[22]    G. M. Sessler, J. E. West, *J. Appl. Phys.* **1971**, *43*, 922.
[23]    K. S. Novoselov, A. K. Geim, S. V. Morozov, D. Jiang, Y. Zhang, S. V. Dubonos, I. V. Grigorieva, A. A. Firsov, *Science* **2004**, *306*, 666.
[24]    Y. Zhang, Y.-W. Tan, H. L. Stormer, P. Kim, *Nature* **2005**, *438*, 201.
[25]    T. Ando, A. B. Fowler, F. Stern, *Rev. Mod. Phys.* **1982**, *54*, 437.
[26]    K. Nomura, A. H. MacDonald, *Phys. Rev. Lett.* **2007**, *98*, 076602.
[27]    N. M. R. Peres, F. Guinea, A. H. Castro Neto, *Phys. Rev. B* **2006**, *73*, 125411.
[28]    E. H. Hwang, S. Adam, S. Das Sarma, *Phys. Rev. Lett.* **2007**, *98*, 186806.
[29]    V. V. Cheianov, V. I. Fal'ko, *Phys. Rev. B* **2006**, *74*, 041403.
[30]    M. I. Katsnelson, K. S. Novoselov, A. K. Geim, *Nat. Phys.* **2006**, *2*, 620.
[31]    V. V. Cheianov, V. I. Fal'ko, B. L. Altshuler, *Science* **2007**, *315*, 1252.